\documentstyle[epsf]{article}

\makeatletter
\def\appendix{\par\clearpage
  \setcounter{section}{0}
  \setcounter{subsection}{0}
  \@addtoreset{equation}{section}
  \def\@sectname{Appendix~}
  \def\theequation{\thesection.\arabic{equation}}
  \def\thesection{\Alph{section}}}
\makeatother

\newcommand{\beq}{ \begin{equation} }
\newcommand{\eeq}{ \end{equation} }
\newcommand{\bea}{ \begin{eqnarray} }
\newcommand{\eea}{ \end{eqnarray} }
\newcommand{\be}{ \beta }
\newcommand{\f}{ \frac }
\newcommand{\de}{ \partial }

\newcommand{\simst}{ \:_\sim^{_<}\, }

\begin{document}
\thispagestyle{empty}
\parskip=12pt
\raggedbottom

\def\mytoday#1{{ } \ifcase\month \or
 January\or February\or March\or April\or May\or June\or
 July\or August\or September\or October\or November\or December\fi
%\space\number\day ,
 \space \number\year}
\noindent
\hspace*{9cm} BUTP--97/28\\
\vspace*{1cm}
\begin{center}

{\LARGE Instanton classical solutions \\ of SU(3) fixed point actions 
on open lattices\footnote{Work supported by Fondazione ``A.~Della
  Riccia''-Italy and INFN-Italy.}  } 

\vspace{1cm}

Federico Farchioni and Alessandro Papa \\
Institute for Theoretical Physics \\ 
University of Bern \\
Sidlerstrasse 5, CH--3012 Bern, Switzerland

\vspace{0.5cm}

\mytoday \\ \vspace*{0.5cm}

\nopagebreak[4]

\begin{abstract}
We construct instanton-like classical solutions of the fixed point
action of a suitable renormalization group transformation
for the SU(3) lattice gauge theory. The problem of the non-existence
of one-instantons on a lattice with periodic boundary conditions is
circumvented by working on open lattices.
We consider instanton solutions for values of the size (0.6-1.9 in lattice
units) which are relevant when studying the SU(3) topology  on coarse lattices
using fixed point actions. We show how these instanton configurations on open
lattices can be taken into account when determining a few-couplings
parametrization of the fixed point action.
\end{abstract}

\end{center}


\vfill

\eject

\section{\bf Introduction}
\label{sec:intro}

Fixed point (FP) actions~\cite{HN94} are classically perfect lattice actions,
in the sense that their spectral properties are free of cut-off effects at the
classical level. Moreover, they possess scale-invariant instanton solutions
up to a minimum size of the order of one lattice spacing~\cite{HN94,BBHN96}.
The
latter property, in conjunction with a suitable definition of the topological
charge operator, allows a theoretically sound approach to the topology on the
lattice~\cite{BBHN96,DFP97,Nie97}.
In particular, since the FP action takes the continuum value on the classical
solutions, the instanton size distribution in the canonical
ensemble is just that of the continuum up the cut-off scale. In the case of
the Wilson action, on the contrary, the instanton action is subject to a
very fast decrease for decreasing size, and, as a consequence, short-ranged
topological fluctuations are over-produced in the thermal ensemble.
These cut-off-scale configurations, which are in fact lattice artifacts,
hamper the accurate evaluation of the topological quantities (e.g. the
topological susceptibility $\chi$) by Monte Carlo simulations, since their
topological charge cannot be unambiguously defined on the lattice.

For a given converging renormalization group (RG) transformation,
the FP action of any lattice configuration is well-defined and can be
determined numerically with arbitrary accuracy by multigrid
minimization~\cite{HN94}. In Monte Carlo simulations of SU(3)
lattice gauge theory, however, due to computing time limitations
only few-couplings parametrizations of the FP
action can be used. These parametrizations are obtained by a fit procedure on
the values of the FP action of a representative set of thermally generated
lattice configurations~\cite{DHHN95b}. Even if for a parametrized form
the theoretical properties of the FP action hold only in approximate sense,
parametrized FP actions have shown practically no cut-off dependence in Monte
Carlo simulations for spectral quantities like the string tension and the free
energy density even on very coarse lattices~\cite{DHHN95b,Pap96}.

Topology deserves a separate discussion, since the implementation in a
pa\-ra\-me\-tri\-zed ver\-sion of the properties of the (exact) FP action
concerning topological classical solutions introduces additional
difficulties. Indeed, classical solutions never appear among the thermal
configurations of the equilibrium ensemble, which are the kind of
configurations considered in the above mentioned parametrizations; if one
wants to include the information of the scale-invariance of the FP action in
the parametrization procedure, classical solutions of the FP action must be
``artificially'' constructed and included in the set of configurations
considered in the fit. In consideration of this, it is not a surprise that the
first pa\-ra\-me\-tri\-za\-tions adopted in Monte Carlo simulations, e.g. the
``type IIIa'' action proposed in Ref.~\cite{BN96}, which perform amazingly
well in the case of spectral properties, show (as we will see in the following)
practically no special properties for topology when compared to the standard
discretization, i.e. the Wilson action.

It has been observed~\cite{DHHN95b,Pap96} that the parametrized FP actions
exhibit physical scaling already for lattices with spacing $a \simst 1$
fm. Since this is just the scale of lengths where the instanton size
distribution of the continuum reaches its maximum~\cite{DGS97}, the typical
instanton size coming into play for such lattices is of the order of the
lattice spacing. The consequence is that, if the topological properties
of the continuum theory have to be fulfilled in this precocious scaling
region, one should concentrate on instantons of size in lattice units
$\hat{\rho} \sim 1$\footnote{Here and in the following, the hat on a
  dimensionful quantity indicates that it is expressed in lattice
  units.}.

The aim of this work is to construct, for this relevant region of size,
instanton-like classical solutions of the FP action of the type III RG
transformation~\cite{BN96}. These lattice configurations can be subsequently
included in the fit procedure for the determination of a new action
parametrization. This parametrized FP action is expected to well reproduce the
FP action of both thermal lattice configurations and classical
solutions\footnote{One could deduce that such parametrized FP action in
  addition should mimic the exact FP action even in the case of a topological
  classical solution with superimposed quantum fluctuations.}. The choice of
the type III RG transformation is motivated by its property to bring to a FP
action with a quite short interaction range~\cite{BN96} and therefore suitable
to be parametrized in terms of a few loops extended over 1-2 lattice spacings.

The main obstacle to this project is represented by the
non-existence of one-instantons on a lattice with periodic
boundary conditions. An effective approach in this context, still with
periodic b.c., has been proposed in Ref.~\cite{DHZ96a}, allowing
the construction of a few-couplings parametrization of the FP action
for SU(2)~\cite{DHZ96b,DeGHaKo97}. Here, we adopt an
alternative procedure, which avoids to make use of b.c. at all, i.e. working
on open lattices.
If on one side this approach solves the problem at the root,
on the other it introduces many new technical subtleties. In this paper we
show how all these technicalities can be overcome,
allowing the construction of
instanton classical solutions on open lattices and their
insertion in the parametrization procedure in a consistent framework.

An elegant way to accommodate a single instanton
on the lattice is also to use twisted b.c.~\cite{twist}.
Except for the lack of an analytical form for the instanton classical
solution and for the need to find a proper definition of the instanton size,
the procedure followed in this paper could be rephrased
without problems for the case of twisted b.c..

The remaining of the paper is organized as follows: in Section~\ref{sec:fp},
we briefly review the definition of FP action and discuss some consequences
of its scale-invariance in connection with topology; in
Section~\ref{sec:block}, we present our procedure to construct instanton
classical solutions of the FP action of the type III RG transformation on open
lattices; in Section~\ref{sec:par}, we show how these instanton
configurations can be involved in the parametrization of the FP action;
in Section~\ref{sec:concl}, we present our conclusions.

\section{Fixed point actions and topology}
\label{sec:fp}

The partition function of a SU(N) lattice gauge theory is
\beq
Z = \int DU \: e^{-\beta {\cal A}(U)} \;\;\; ,
\label{eq:part}
\eeq
where $\be {\cal A}(U)$ is any lattice regularization of the continuum action
expressed in terms of products of link variables $U_\mu(n)\in$ SU(N) along
arbitrary closed loops. Denoting with $\be$, $c_2, c_3, \ldots$
the couplings contained in $\be {\cal A}(U)$, the action can be represented
as a point in the infinite dimensional space of the couplings.

A RG transformation in this space of the couplings with scale factor 2
can be defined in the following way:
\beq
e^{- \be^{\prime}{\cal A}^{\prime}(V)} = \int DU\: e^{-\be [\: {\cal
    A}(U)+T(U,V)\:]}\;\;\;;
\label{eq:RG}
\eeq
here $\{V\}$ denotes the coarse configuration living on the lattice with
spacing $a^\prime=2\, a$, whose links $V_\mu(n_{\rm B})$ are related to a
local average of the original link variables $U_\mu(n)$; $T(U,V)$ is the
blocking
kernel which defines this
average, normalized in order to keep the partition function invariant under
the transformation. The explicit form of $T(U,V)$ is
\beq
T(U,V) = \sum_{n_{\rm B},\mu}\left(\;{\cal N}_\mu(n_{\rm
    B})-\f{\kappa}{N}{\rm
    Re\;Tr} [\:V_\mu(n_{\rm B})\:Q_\mu^\dag(n_{\rm
    B})\:]\;\right) \;\;\;,
\label{eq:T}
\eeq
where $Q_\mu(n_{\rm B})$ is a $N\times N$ matrix which represents some
average of paths of fine link variables $U_\mu(n)$ connecting
the sites $ 2n_{\rm B}$ and $2n_{\rm B}+2\hat\mu$; ${\cal N}_\mu(n_{\rm B})$
guarantees the normalization, while $\kappa$ is a free parameter
adjustable for numerical optimization.
For the type III RG transformation, $Q_\mu(n_{\rm B})$ is given by the
product of the fuzzy links ${\mathsf{W}}_\mu(2n_{\rm B})$
and ${\mathsf{W}}_\mu(2n_{\rm B}+\hat\mu)$, each of them being a weighted sum
of paths of fine links living on the $1^4$ hypercubes which contain also
$U_\mu(2n_{\rm B})$ and $U_\mu(2n_{\rm B}+\hat\mu)$, respectively (see
Ref.~\cite{BN96} for the details, and for the values of $\kappa$ and of
the other free parameters of the RG kernel).

On the critical surface $\be=\infty$, Eq.~(\ref{eq:RG}) leads to the saddle
point equation
\beq
{\cal A}^\prime (V) = \min_{\{U\}} \; [ \, {\cal A}(U)+T(U,V) \, ] \;\;\; ;
\eeq
the FP point of this transformation is therefore
\beq
{\cal A}^{\rm FP} (V) = \min_{\{U\}} \; [ \, {\cal A}^{\rm FP}(U)+T(U,V) \, ]
\;\;\; .
\label{eq:FP}
\eeq
In the limit $\beta\rightarrow\infty$, ${\cal N}_\mu(n_{\rm B})$
is given by
\beq
{\cal N}_\mu(n_{\rm B}) = \f{\kappa}{N}
\max_{W\in{\rm SU(3)}}{\rm Re\;Tr} [\:W\:Q_\mu^\dag(n_{\rm B})\:]\;\;\;.
\label{eq:N}
\eeq

A blocking transformation on the link variables $U_\mu(n)$ can be defined,
where the blocked link $V_\mu(n_{\rm B})$ is obtained by projecting on SU(3)
the average $Q_\mu(n_{\rm B})$. In other words, $V_\mu(n_{\rm B})$ is given
by the SU(3) matrix $W$ maximizing the quantity
\beq
{\rm Re\;Tr}[\:W\: Q_\mu^\dag(n_{\rm B})\:]\;\;\;;
\label{eq:block}
\eeq
it is evident from Eqs.~(\ref{eq:T}) and (\ref{eq:N}) that for a given fine
configuration $\{U\}$, the corresponding blocked configuration $\{V\}$ is the
solution of the equation $T(U,V)\:=\:0$ at $\beta=\infty$.

An important consequence of Eq.~(\ref{eq:FP}) is the following: if a
configuration $\{V\}$ satisfies the FP classical equations and is a local
minimum
of ${\cal A}^{\rm FP} (V)$, then the fine configuration $\{U(V)\}$
minimizing the r.h.s. of Eq.~(\ref{eq:FP}) satisfies the FP classical
equations as well and the value of the action remains
unchanged~\cite{HN94,DHHN95a}:
\beq
{\cal A}^{\rm FP}(V)\:=\:{\cal A}^{\rm FP} (U(V))\;\;\;.
\eeq
Moreover, since $T(U(V),V)=0$, $\{V\}$ is just the configuration obtained by
blocking the fine configuration $\{U(V)\}$. For this reason, it can be said
that $\{U(V)\}$ represents the ``inverse-blocked'' configuration of
$\{V\}$. If $\{V\}$ is an instanton-like classical solution of the FP action
with size $\hat\rho$ in coarse lattice units, the configuration $\{U(V)\}$ is
also an instanton-like classical solution with the same action and size
$\hat\rho^\prime=2\hat\rho$ in fine lattice units\footnote{The factor two
  comes from the fact that the size in physical units $\rho$ does not change
  under blocking.}: the action is scale-invariant. By iteration one can
conclude that this scale-invariant instanton action is just that of the
continuum, ${\cal A}_{\rm inst}=4\pi^2/N$. The theorem does not hold in the
reversed direction: starting from a classical solution, it is not guaranteed
that the blocked configuration is still a classical solution with halved size
in the blocked lattice units. In fact, there is a critical instanton size in
lattice units $\hat\rho_c$, below which no instanton classical solutions can
exist on the lattice even with the FP action. This critical size turns
out to be typically of the order of 1 in lattice units~\cite{BBHN96}.
In order to check that the blocked configuration of an instanton solution is
still a solution, it is necessary in principle to verify that the
inverse-blocking restores the original configuration.

On the basis of the above considerations, a procedure to construct instanton
classical solutions of the FP action of a given RG transformation up to a size
as small as the critical one can be outlined as follows~\cite{BBHN96}: on a
very fine lattice a continuum instanton solution with a large size
$\hat\rho_0$ is na\"{\i}vely discretized; for large enough $\hat\rho_0$, the
lattice configuration so obtained approximates the corresponding classical
solution of the FP action with high accuracy.
Then, starting from this configuration, a certain number of blocking
transformations is performed in order that the size of the blocked instanton
after the last step is of order 1 in units of the final blocked lattice. As
stated in the Introduction, this is indeed the relevant region of size when
simulating with FP actions. That the instanton is not destroyed by the
blocking transformation can be checked by inverse-blocking, as explained above.

\section{Constructing fixed point classical solutions on open lattices}
\label{sec:block}

In this Section we put into practice the procedure outlined at the end of the
previous Section. The first step is the discretization of a continuum instanton
solution on the lattice. The size of the instanton in lattice units has to be
large enough to have a smooth lattice configuration. If one starts from a
finite lattice with periodic b.c., a difficulty arises since one-instanton
solutions with such b.c. do not exist in the continuum. One possibility is
to take the infinite volume one-instanton solution in the continuum, to cut it
on a finite lattice and to force periodicity. This introduces finite volume
effects in the action which
scale with the lattice size $L$~\cite{DHZ96b}. This severe volume dependence
can be damped to order $1/L^3$ by performing a singular gauge transformation
on the instanton configuration at infinite volume, i.e. before cutting it to
the finite lattice~\cite{PT89,DHZ96b}. The lattice configuration obtained in
this way is the starting configuration of the blocking procedure. This is a
good approximation of a FP classical solution in local sense except in the
vicinity of the border, where a systematic deviation comes into play.

Our approach is radically different since it avoids to impose periodic
b.c. in any step of the full procedure. How this can be done is clear if one
considers the structure of the blocking kernel $T(U,V)$ (Eq.~(\ref{eq:T})): the
blocked link $V_\mu(n_{\rm B})$ is affected only by the fine links entering
the definition of $Q_\mu(n_{\rm B})$, which, for the type III RG
transformation,
live on the $1^4$ hypercubes of the fine lattice containing the fine links
$U_\mu(2 n_{\rm B})$ and $U_\mu(2 n_{\rm B}+\hat\mu)$. Consequently, in order
to perform a blocking transformation on a lattice $\{n: \; n_\mu=-\hat
L/2, \ldots, \hat L/2 -1, \; \mu=1,\ldots,4\}$ without imposing any b.c.,
it is necessary to know the infinite volume configuration on the
larger lattice $\{n: \; n_\mu=-(\hat L+2)/2, \ldots, (\hat L+2)/2 -1, \;
\mu=1,\ldots,4\}$. In this way, the blocking transformation is locally
equivalent to that in the infinite volume. Of course, the blocked
configuration (as well as the starting one) is known only on a finite
sub-lattice of an ideal infinite lattice. In
the following, we will refer to sub-lattices embedded in an infinite lattice
without any prescription for their boundary as to ``open''
lattices\footnote{Open lattices in the context of SU(2) topology were used
  already in Ref.~\cite{NN95}.}. Given an
open lattice $\Lambda$ with size $\hat L$ and a configuration $\{U\}$ living
on the infinite lattice embedding $\Lambda$, we define $\{U^{\rm int}\}$ as the
set of the links $U_\mu(n)$ of the configuration $\{U\}$ originating from any
site $n$ which belongs to $\Lambda$ along both positive and negative
directions. All the other links of $\{U\}$ will be denoted by $\{U^{\rm
ext}\}$.

\subsection{Lattice action on open lattices}
\label{subsec:act}

Our way of proceeding poses the problem of assigning a lattice action to a
configuration living on an open lattice. Even once that a lattice action
regularization is fixed, there is the further arbitrariness in prescribing
which loops are to be considered internal and which external to the open
lattice.

We consider lattice action regularizations of the form
\beq
{\cal A} = \f{1}{N}\sum_{C, \ i\geq1} c_i(C)[N-{\rm Re\;Tr}(U_C)]^i \;\;\;,
\label{eq:par}
\eeq
where $C$ denotes any closed path, $U_C$ stands for the product of the link
variables along the path $C$ and $c_i(C)$ is the coupling associated to the
$i$-th power for the loop $C$ (the couplings $c_1(C)$ must satisfy the
normalization condition which ensures the correct continuum limit, see
Appendix). For an open lattice $\Lambda = \{n: \; n_\mu=-\hat
L/2, \ldots, \hat L/2 -1, \; \mu=1,\ldots,4\}$,
after a definite choice of the topologies of closed paths $C$ (typically
living on a $1^4$ or $2^4$ hypercube), we define the lattice action $ {\cal
  A}_{\rm int} = \sum_{n\in\Lambda} {\cal A}(n) $ through the clover-averaged
action density~\cite{GGSV94}
\beq
{\cal A}(n) = \f{1}{N} \sum_{C\ni n, i\geq1} c_i(C) \f{[N-{\rm
    Re\;Tr}(U_C)]^i}{{\rm perimeter}(C)} \;\;\;.
\label{eq:act_dens}
\eeq
Here the most important point is that only those loops $C$ which contain the
site $n\in\Lambda$ are summed over.
The action density ${\cal A}(n)$ has the right continuum limit and when summed
on the sites of the infinite lattice reproduces the lattice
action~(\ref{eq:par}). It should be noted that the
definition~(\ref{eq:act_dens}) involves also links which are external to the
lattice $\Lambda$, but belong to closed paths $C$ containing internal
sites. Of course, after summing ${\cal A}(n)$ over the sites of the open
lattice
$\Lambda$, the weights $1/{\rm perimeter}(C)$ for any given loop topology $C$
sum up to 1 for the internal loops, while for the loops on the border they sum
to a fractional value.

If an instanton configuration is known only for the links contained in an open
lattice, the simplest guess for its full infinite volume action is obtained by
adding to ${\cal A}_{\rm int}$ the action density of the continuum integrated
on the external volume\footnote{Here we adopt the convention of anti-Hermitian
  gauge fields $A_\mu(x)$ (see Appendix).}:
\beq
{\cal A} = {\cal A}_{\rm int} + {\cal A}_{\rm ext} \simeq {\cal A}_{\rm int} +
\int_{{\cal R}^4 \setminus \Lambda}
\: d^4\, x \left[ \; -\f{1}{2} \; {\rm Tr} (\: F^{\rm (inst)}_{\mu\nu}(x)
  \:)^2 \;\right] \;\;\; ,
\label{eq:act_inst}
\eeq
where ${\cal R}^4 \setminus \Lambda$ indicates the volume external to the open
lattice $\Lambda$. Considering the symmetry of the definition of the action
density, Eq.~(\ref{eq:act_dens}), it is natural to take as external volume the
region identified by $x_\mu < -(\hat L +1)a/2 \; , \; x_\mu > (\hat L -1)a/2$,
$\mu=1,\ldots,4$. Eq.~(\ref{eq:act_inst}) could seem a crude
approximation, but as we will show shortly it can be quite accurate if some
conditions on $a/\rho$ and $\rho/L$ are satisfied.

\begin{figure}[htb]
\begin{center}
\leavevmode
\epsfxsize=90mm
\epsfbox{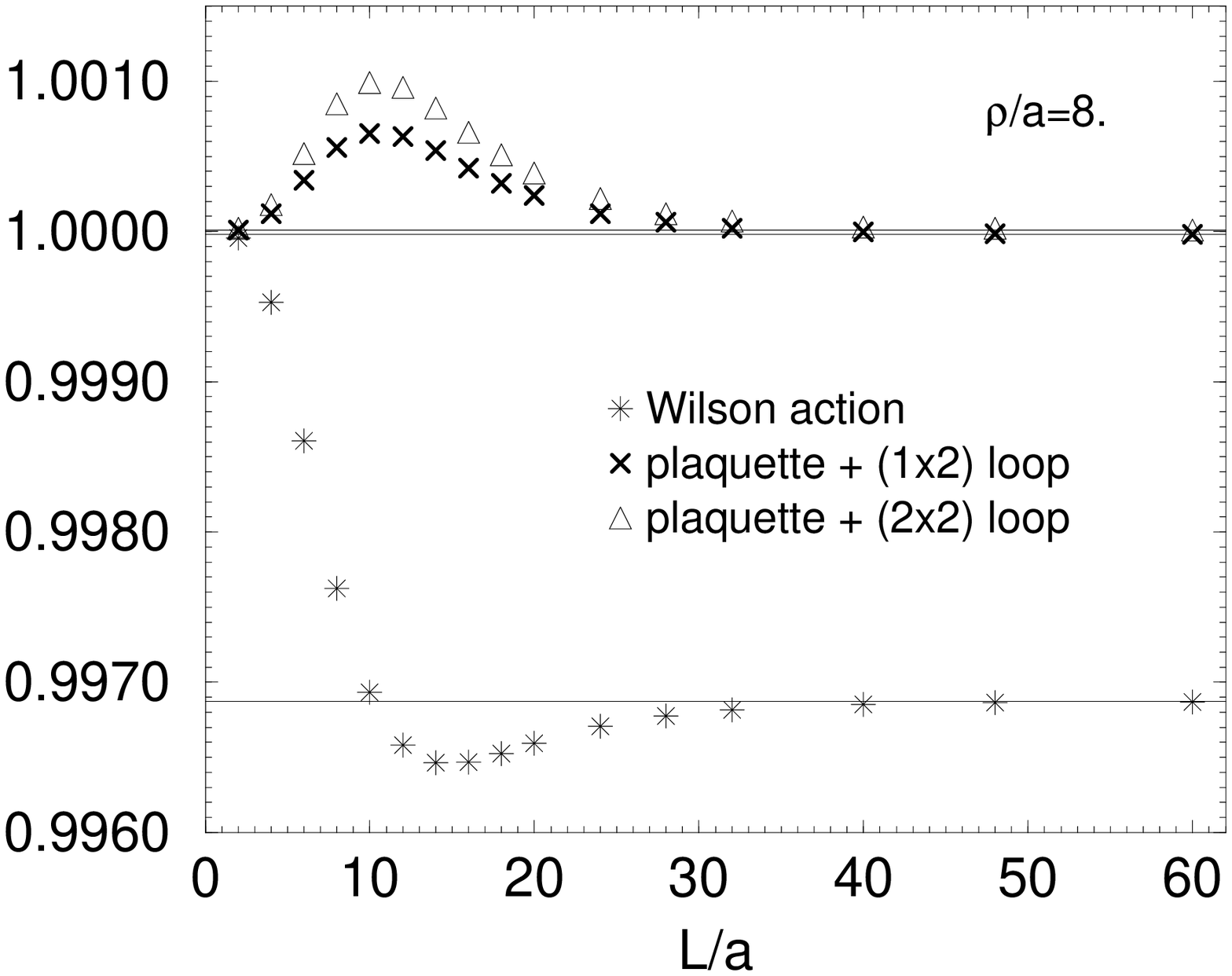}
\caption[]{Estimate through~(\ref{eq:act_inst}) of the infinite volume
  action of a discretized instanton with $\hat{\rho}=8$ known on open lattices
  with size $\hat{L}$. 
  The action is given in units of the continuum value; we consider 
  the Wilson action, the plaquette + $(1\times2)$ loop Symanzik action and
  the plaquette + $(2\times2)$ loop Symanzik action. The solid lines represent
  the expected values of the infinite volume lattice action: 0.99686923,
  0.99998172, 1.00001427, for the three action regularizations respectively.}
\label{fig:voleff_s8}
\end{center}
\end{figure}

\begin{figure}[htb]
\begin{center}
\leavevmode
\epsfxsize=90mm
\epsfbox{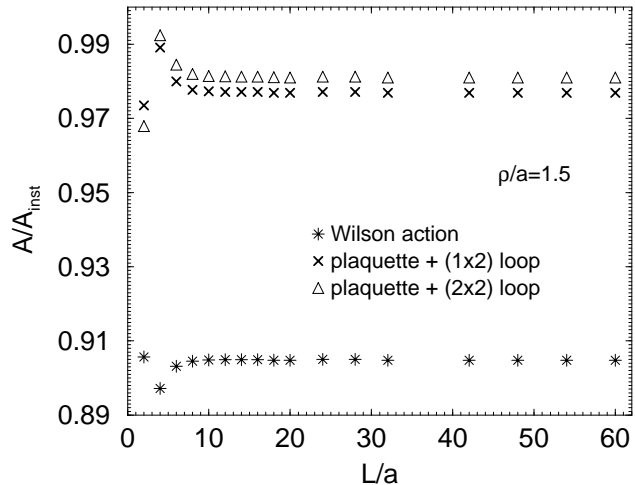}
\caption[]{As in Fig.~\ref{fig:voleff_s8} with $\hat\rho=1.5$. In this case a
  theoretical value for the infinite volume action is not available.}
\label{fig:voleff_s1.5}
\end{center}
\end{figure}

In order to check the accuracy of the Eq.~(\ref{eq:act_inst}), we consider
lattice instanton configurations in SU(2) defined by~\cite{GGSV94}
\beq
U_\mu^{\rm (inst)} (n)={\rm P}\exp \int_0^a A_\mu^{\rm (inst)} (n+s\hat\mu) \:
{\rm d}s \;\;\; ,
\label{eq:Pexp}
\eeq
where $A_\mu^{\rm (inst)} (n)$ are the anti-Hermitian gauge fields for the
continuum instanton (see Appendix for the explicit expression and for
other details needed in the following). The definition~(\ref{eq:Pexp})
corresponds to a lattice instanton obtained by an infinite number of blocking
steps from the continuum according to the simple blocking
transformation\footnote{Unfortunately, this RG transformation does not
  converge to a FP in $d=4$.} $V_\mu(n_{\rm B})=U_\mu(2n_{\rm B})U_\mu(2n_{\rm
  B}+\hat\mu)$. Expanding in powers of $a^2$ the action
density~(\ref{eq:act_dens}) of the configuration $U_\mu^{\rm (inst)}(n)$ given
in~(\ref{eq:Pexp}), an infinite series is obtained, where the term $\sim
a^{4+2n}$ is a linear combination of the
continuum gauge-invariant operators with na\"{\i}ve dimension $4+2n$.
The leading term of this expansion is just $-1/2\: {\rm Tr}
(\:F_{\mu\nu}^{\rm (inst)}(x)\:)^2 a^4$, i.e. the action density on
the continuum which appears in the integral at the
r.h.s. of~(\ref{eq:act_inst}). For a continuum instanton with size $\rho$,
this integral depends only on $\rho/L$ and is proportional to
$(\rho/L)^4$ for small $\rho/L$. The first correction to the
estimate~(\ref{eq:act_inst}) is due to the term $\sim a^6$ in the expansion of
the action density~$(\ref{eq:act_dens})$ and is proportional to
$(a/\rho)^2(\rho/L)^6$ (see Appendix). Higher dimension
operators give contributions proportional to $(a/\rho)^{2n}(\rho/L)^{4+2n}$,
$n=2,3,\ldots$ . The corrections are therefore under control when
$a/\rho\simst1$ and $\rho/L$ is sufficiently small. In
Fig.~\ref{fig:voleff_s8} we report the estimate~(\ref{eq:act_inst}) of the
action of a lattice instanton with size $\hat\rho=8$ for different values of
the size $\hat L$ of the open lattice; we consider three different
lattice actions, the Wilson action, the plaquette + $(1\times2)$ loop
tree-level Symanzik improved action and the plaquette + $(2\times2)$
loop tree-level Symanzik improved action\footnote{In the case of tree-level
  Symanzik improved actions, the
  corrections to the estimate~(\ref{eq:act_inst}) are more suppressed, since
  they start from $(a/\rho)^4 (\rho/L)^8$.}. We observe that the discrepancy
between the estimate~(\ref{eq:act_inst}) and the theoretical action value
goes down rapidly for increasing $\hat L$. The theoretical value
is obtained by the expansion of the action in powers of $a/\rho$ (see
Appendix) up to the order 4; for $\hat\rho=8$
this is an excellent approximation. In Fig.~\ref{fig:voleff_s1.5} we perform
the same comparison in the less favorable case $\hat\rho=1.5$, for which the
expansion of action in $a/\rho$ is not enough accurate. We observe in
this case a fast convergence to a plateau, corresponding to the exact
infinite volume action. This value is approximated within some per mill
already for $\hat L\geq6$.

\subsection{The blocking procedure}
\label{subsec:block}

The next step in the procedure outlined at the end of Section~\ref{sec:fp}
consists in performing repeatedly blocking transformations on smooth instanton
configurations until their size is of the order of 1 in units of the
final blocked lattice. \newline
We denote the instanton configuration on the finest lattice, at the starting
point of the blocking procedure, by $\{U_0\}$; the starting instanton size in
lattice units is $\hat\rho_0$. The instanton configuration after
$k$ blocking steps is indicated with $\{U_k\}$; its size in units of the
spacing of the $k$-blocked lattice is $\hat\rho_k = \hat\rho_0/2^k$.

As we have seen in the previous Subsection, for instantons with size of the
order of 1 lattice spacing, a $6^4$ open lattice allows an accurate
determination of the lattice instanton action at infinite volume through
Eq.~(\ref{eq:act_inst}). In order to
calculate on an open lattice the contribution ${\cal A}_{\rm int}$
to the total action ${\cal A}$ according to Eq.~(\ref{eq:act_inst}), we must
be able to calculate also those loops entering the
definition~(\ref{eq:act_dens})
which lie on the border of the lattice and involve external links. This
amounts to know the configuration on a lattice larger than the open lattice
considered.
For our purposes, in order to calculate ${\cal A}_{\rm int}$ on open lattices
$6^4$, we need to know the configurations on lattices $10^4$; this is
sufficient,
since we are interested in loops which do not extend beyond the $2^4$
hypercube. We have therefore to organize the blocking procedure in order that
the lattice size after the last blocking step is 10. \newline
Now, if there were definite b.c., a $10^4$ lattice could
be obtained by blocking from a $20^4$ lattice. In our approach, where
no b.c. are imposed, a $10^4$ lattice can be obtained by blocking from
$22^4$ lattice, according to what discussed at the beginning of
Section~\ref{sec:block}. In the case of three blocking steps, which will turn
out to be enough for our purposes, the sequence of lattice size is
$94\rightarrow 46 \rightarrow 22 \rightarrow 10\,$\footnote{We
overcome the memory problems associated to such large lattices by allocating
smaller lattices and exploiting the symmetries of instanton configurations,
which we have verified to be conserved by the blocking procedure.}.

The instanton configuration on the finest lattice is obtained by
discretization according to~(\ref{eq:Pexp}). Since the instanton size halves
after each of the three blocking steps, the typical $\hat\rho_0$ on the finest
configuration has to be around 8 lattice units. For such instanton sizes any
lattice action gives the same value within some per mill (for $\hat\rho=8$,
using the $a/\rho$ expansion up to order $(a/\rho)^4$, one obtains 0.99686923
for the Wilson action, 1.00001427 for the tree-level Symanzik
improved action with the $(2\times2)$ loop, 0.99998172 for the tree-level
Symanzik improved action with the $(1\times2)$ loop). We deduce that such
configurations are a good approximation of the continuum classical
solutions. It must be checked, however, that they remain solutions of the FP
classical equations of motion with the same approximation even after
three blocking steps (we recall the discussion in Section~\ref{sec:fp}). This
can be done by calculating the FP action of the final configuration $\{U_3\}$
and verifying that it reproduces the continuum action. The FP action of
$\{U_3\}$ is formally given by Eq.~(\ref{eq:FP}), which in this case reads
\beq
{\cal A}^{\rm FP} (U_3) = \min_{\{U\}} \; [ \, {\cal A}^{\rm FP}(U)+T(U,U_3)
\, ] \;\;\;.
\label{eq:FPappl}
\eeq
One should iterate this equation until the relevant configurations on the
finest lattice are so smooth that any lattice action can be used in the
minimization. In practice, for four-dimensional non-Abelian gauge theories,
memory and time limitations prevent from performing more than one
step. Fortunately, after one step of inverse-blocking the minimizing
configuration $\{U_{\rm min}\}$ is already so smooth (the action
lowers typically by a factor 30-40~\cite{Nie97}) that ${\cal A}^{\rm FP}(U)$
can be well approximated by a tree-level Symanzik improved action; in this
context we use the plaquette + $(1\times2)$ loop Symanzik
action~\cite{DHZ96b}. \newline Another point to check is that, being $\{U_3\}$
a FP classical solution, it goes back to $\{U_2\}$ by
inverse-blocking. Rigorously, the full procedure should be carried out on
configurations $\{U_3\}$ and $\{U\}$ living on infinite coarse and fine
lattice, respectively. Since of course we can work
only on finite lattices, it is necessary to introduce some simplifications, to
be justified within our approach with open lattices.\newline
Our procedure is the following: being $\{U_3\}$ given on a $6^4$ lattice,
we search the minimizing configuration $\{U_{\rm min}\}$ by locally updating
only the links $U_\mu(n)$, $\mu=1,\ldots,4$ with $n$ given by
\beq
n=2n_{\rm B}+\sum_\sigma\lambda_\sigma \hat\sigma\;\;\;,\;\;\;\;\;
\lambda_\sigma=0,1\;\;\;, \;\;\;\;\; \sigma=1,\ldots,4 \;\;\; ,
\label{eq:flatt}
\eeq
for any site $n_{\rm B}$ of the $6^4$ coarse lattice. Moreover,
we use as trial starting configuration in the minimization the configuration
$\{U_2\}$ itself. This amounts to fix {\it a priori} $\{U_{\rm min}^{\rm
  ext}\}=\{U_2^{\rm ext}\}$ outside the $12^4$ open lattice whose sites are
given by Eq.~(\ref{eq:flatt}). Since $\{U_2^{\rm ext}\}$ is smooth for the
values of $\rho/L$ under consideration, it is legitimate to estimate ${\cal
  A}_{\rm ext}(U_2)$ with ${\cal A}_{\rm ext}^{\rm (cont)}\,$; moreover, we
assume to be zero the contribution to $T(U,U_3)$ coming from the part of the
summation in Eq.~(\ref{eq:T}) involving sites $n_{\rm B}$ of the coarse
lattice not contained in the open lattice.

\begin{table}[htb]
\setlength{\tabcolsep}{1.03pc}
\caption[]{Minimized and parametrized FP action in units of the continuum
  instanton action of the $\{U_3\}$ configurations for several values of
  $\hat\rho_3$. The configurations $\{U_3\}$ are known on a $6^4$ open lattice
  and their infinite volume action is estimated as explained in the text. The
  couplings of the parametrized FP action are given in
  Table~\ref{tab:new_par}. Column 2 contains the values of $T(U_{\rm
  min},U_3)$ in units of the continuum instanton action resulting from the
  minimization.} 
\centering
\begin{tabular}{cccc}
\hline
$\hat\rho_3$ & $T(U_{\rm min},U_3)/{\cal A}_{\rm inst}$ & 
${\cal A}^{\rm FP}(U_3)/{\cal A}_{\rm inst}$ & ${\cal A}^{\rm FP}_{\rm
  par}(U_3)/{\cal A}_{\rm inst} $\\ 
\hline
 0.1250 & 0.007628 & 0.0255527 & 0.02440536 \\
 0.2500 & 0.018860 & 0.0552252 & 0.05289109 \\
 0.3750 & 0.049999 & 0.1348518 & 0.12942630 \\
 0.5000 & 0.100250 & 0.2797359 & 0.27778473 \\
 0.5625 & 0.004224 & 0.3737279 & 0.38207865 \\
 0.6250 & 0.003092 & 0.9885723 & 0.50021130 \\
 0.6875 & 0.002587 & 0.9887317 & 0.62110525 \\
 0.7500 & 0.002215 & 0.9893495 & 0.73261547 \\
 0.8125 & 0.001924 & 0.9899657 & 0.82569128 \\
 0.8750 & 0.001684 & 0.9903682 & 0.89621836 \\
 1.0000 & 0.001306 & 0.9903730 & 0.97390002 \\
 1.1250 & 0.001013 & 0.9895449 & 0.99372518 \\
 1.2500 & 0.000782 & 0.9882614 & 0.98836493 \\
 1.3750 & 0.000602 & 0.9867958 & 0.97771794 \\
 1.5000 & 0.000464 & 0.9853111 & 0.96955895 \\
 1.6250 & 0.000358 & 0.9839012 & 0.96517944 \\
 1.7500 & 0.000278 & 0.9826250 & 0.96364480 \\
 1.8750 & 0.000218 & 0.9815158 & 0.96376640 \\
\hline
\end{tabular}
\label{tab:min}
\end{table}

Since the finest configurations $\{U_0\}$ are built in SU(2) and since the
blocking is transparent to the color indices, the blocked configurations
$\{U_3\}$ are in SU(2) as well. On this basis, we searched for $\{U_{\rm
  min}\}$ directly in SU(2), verifying in some selected cases that the
minimization in SU(3) brings to the same configuration.

\begin{figure}[htb]
\begin{center}
\leavevmode
\epsfxsize=90mm
\epsfbox{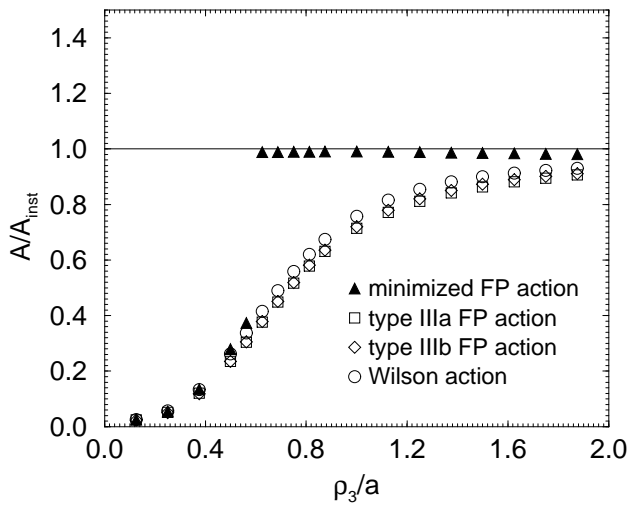}
\caption[]{Minimized FP action (filled triangles), type IIIa FP action
  (squares), type IIIb FP action (diamonds) and Wilson action (circles) of the
  configurations $\{U_3\}$ in units of the continuum instanton action for
  different values of the size $\hat\rho_3$. The configurations $\{U_3\}$ are
  known on a $6^4$ open lattice and their infinite volume action is estimated
  as explained in the text.}   
\label{fig:srho_typeIII}
\end{center}
\end{figure}

Coming to the results, we find that for $\hat\rho_3$ larger than a critical
size $\hat\rho_c\simeq 0.6$, $T(U_{\rm min},U_3)$ is of the order of
$10^{-3}$, showing that $\{U_{\rm min}\}$ is very close to the configuration
$\{U_2\}$ even in the internal region. Moreover, the quantity
\beq
{\cal A}_{\rm int}(U_{\rm min}) + {\cal A}_{\rm ext}^{\rm (cont)} +
T(U_{\rm min},U_3) \;\;\; ,
\label{eq:A+T}
\eeq
that is our estimate for ${\cal A}^{\rm FP}(U_3)$, reproduces within $\sim1\%$
the continuum action for an instanton (see Fig.~\ref{fig:srho_typeIII} and
Table~\ref{tab:min}). The small deviation is quite $\hat\rho_3$-independent
and can be attributed to the fact that the instanton configurations $\{U_0\}$
are not exact instanton classical solutions and/or that the tree-level
Symanzik improved action slightly deviates from the FP action even for the
smooth configurations of the fine lattice involved in the minimization.

For $\hat\rho_3 < \hat\rho_c$ the configurations $\{U_3\}$ are not classical
solutions anymore since ${\cal A}(U_{\rm min}) + T(U_{\rm min},U_3)$ is much
below
the continuum value. These configurations $\{U_3\}$ represent in fact
non-topological quantum fluctuations with range one lattice spacing.

\section{Parametrization of the fixed point action}
\label{sec:par}

Once that we have built on a $6^4$ open lattice a set of configurations
$\{U_3\}$ representing instanton classical solutions of the type III FP
action in the region of size $\hat\rho\sim1$, and have found the
corresponding minimizing configurations $\{U_{\rm min}\}\simeq \{U_2\}$ in the
sense of Eq.~(\ref{eq:FP}), we consider in this Section the problem of how
these
configurations can be used to determine a few-couplings parametrization of the
FP action which keeps track, with a good approximation, of the scale
properties of the exact FP action.

\begin{table}[tb]
\setlength{\tabcolsep}{.68pc}
\caption[]{List of the loops living on a $2^4$ hypercube: they are classified 
  according to the 28 different topologies. For each kind of loop, a
  representative orientation, the multiplicity and the 
  contribution to the operators ${\cal O}_0$, ${\cal O}_1$, ${\cal O}_2$ and
  ${\cal O}_3$ (see Appendix) are given.}   
\centering
\begin{tabular}{ccccccc}
\hline
loop & path & multiplicity & $r_0$ & $r_1$ & $r_2$ & $r_3$ \\ 
\hline
  1  &  x, y, -x, -y                &   6 &   1 &     1 &    0 &    0 \\  
  2  &  x, y, y, -x, -y, -y         &  18 &   8 &    20 &    0 &    0 \\
  3  &  x, y, z, -y, -x, -z         &  72 &  16 &     4 &    0 &   12 \\
  4  &  x, y, z, -x, -y, -z         &  24 &   8 &   --4 &    4 &    4 \\
  5  &  x,  y, x, y, -x, -y, -x, -y &  24 &   8 &    32 &    0 &    0 \\
  6  &  x, y, x, y, -x, -x, -y, -y  &  48 &  36 &   132 &    0 &    0 \\
  7  &  x, y, x, z, -y, -x, -x, -z  & 384 & 192 &    48 &   96 &  144 \\
  8  &  x, y, x, z, -x, -z, -x, -y  & 192 &  32 &     8 &    0 &   24 \\
  9  &  x, y, x, z, -x, -y, -x, -z  &  96 &  40 &    16 &   24 &   24 \\
 10  &  x, y, x, z, -x, -x, -z, -y  & 384 & 160 &   256 &    0 &   96 \\
 11  &  x, y, x, -y, -x, y, -x, -y  &  24 &   0 &  --12 &    0 &    0 \\
 12  &  x, y, y, x, -y, -x, -x, -y  &  24 &   0 &  --24 &    0 &    0 \\
 13  &  x, y, y, z, -y, -z, -x, -y  & 192 &  32 &    56 &    0 & --24 \\
 14  &  x, y, y, z, -y, -y, -x, -z  & 192 &  80 &    80 &    0 &   96 \\
 15  &  x, y, y, z, -y, -x, -z, -y  & 192 &  48 &    24 & --24 &   72 \\  
 16  &  x, y, y, z, -x, -y, -y, -z  &  96 &  72 &    24 &   48 &   48 \\
 17  &  x, y, y, -x, z, -y, -y, -z  &  96 &  64 &   112 &    0 &   48 \\
 18  &  x, y, y, -x, -x, -y, -y, x  &  12 &  16 &    64 &    0 &    0 \\
 19  &  x, y, z, t, -z, -t, -x, -y  & 192 &  32 &    32 &    0 &    0 \\
 20  &  x, y, z, t, -z, -y, -x, -t  & 192 &  48 &  --24 &    0 &   72 \\
 21  &  x, y, z, t, -z, -x, -t, -y  & 384 &  96 &     0 &    0 &   96 \\
 22  &  x, y, z, t, -z, -x, -y, -t  & 384 & 128 & --112 &   48 &  144 \\
 23  &  x, y, z, t, -y, -x, -t, -z  &  96 &  32 &  --16 &    0 &   48 \\
 24  &  x, y, z, t, -y, -x, -z, -t  & 192 &  80 & --112 &   48 &   96 \\
 25  &  x, y, z, t, -x, -y, -z, -t  &  48 &  24 &  --48 &   24 &   24 \\ 
 26  &  x, y, z, -y, -z, y, -x, -y  &  96 &  16 &    28 &    0 & --12 \\ 
 27  &  x, y, z, -y, -x, y, -z, -y  &  48 &   4 &    16 & --12 &   12 \\
 28  &  x, y, -x, -y, x, y, -x, -y  &  12 &   4 &     4 &    0 &    0 \\
\hline
\end{tabular}
\label{tab:loops}
\end{table}

Before proceeding in this program, we check how the previous
parametrizations of the type III FP action proposed in Ref.~\cite{BN96} behave
on the instanton configurations $\{U_3\}$. These parametrizations, called type
IIIa and type IIIb, were found by fitting the exact FP action, determined by
numerical minimization, on a set of $\sim 500$ thermal Monte Carlo
configurations generated on a $2^4$ lattice in the range $\beta_{\rm
  Wilson}=5.1-50$. Both parametrizations are of the form~(\ref{eq:par}) and
involve the plaquette and the twisted perimeter-6 loop (entries no.~1 and
4 in Table~\ref{tab:loops}, respectively) with $i\leq4$~\cite{BN96}. In the
type IIIb parametrization it was imposed that the couplings satisfy the
on-shell tree-level Symanzik condition at the quadratic order in the gauge
fields.

In Fig.~\ref{fig:srho_typeIII} we present the behavior of the type IIIa and
type IIIb actions on the configurations $\{U_3\}$ in comparison with the
Wilson action. We can see that they fail to reproduce the scale-invariant
shape of the minimized FP action to the same extent as the Wilson action. This
is not a surprise since these parametrizations were not required to well
reproduce the FP action on classical solutions, but on thermal
configurations. We see also that the tree-level Symanzik improvement at the
quadratic level in the gauge fields has no effect on the action of instanton
classical solutions.
For the values of $\beta$ where the FP actions are expected to display
physical scaling~\cite{DHHN95b,Pap96}, corresponding to  $a \simst 1$ fm,
the relevant region in the instanton size is $\hat{\rho}\sim 1$~\cite{DGS97};
in these conditions, the behavior exhibited in Fig.~\ref{fig:srho_typeIII}
by the type IIIa and type IIIb actions could be responsible for distortions
of the lattice signal when computing topological quantities~\cite{DeGHaKo97}.
Indeed, the small-size ($\hat{\rho}\sim 1$) topological configurations,
having an action lower than the continuum instanton action,
are over-produced in the thermal ensemble and their
contribution to the total topological signal is over-estimated,
when a non-zero topological charge is assigned to them~\cite{BBHN96,DFP97}.

Parametrizations of the FP action which approximately possess the
scale-invariance properties of the exact FP action can be built by taking into
account the additional information coming from the classical
solutions~\cite{BBHN96}. In the following we explain in some detail how this
has been done in the present case.

The fit procedure starts by fixing the form of the parametrization of the FP
action, Eq.~(\ref{eq:par}) in our case, and, accordingly, a finite number of
loop topologies and powers. We have restricted ourselves to all the
loops living on the $2^4$ hypercube (see Table~\ref{tab:loops}) and to four
powers of the trace. \newline
The first condition of the fit is that the FP equation~(\ref{eq:FP}) holds
also at a pa\-ra\-me\-tri\-zed level
\beq
{\cal A}_{\rm par}^{\rm FP} (V) = {\cal A}_{\rm par}^{\rm FP}(U_{\rm
  min})+T(U_{\rm min},V)\;\;\;.
\eeq
In the framework of the fit procedure this means the minimization of the
quantity
\beq
Q_1(c_\alpha) = \sum_{\{V\}} \left[ \sum_\alpha c_\alpha {\cal T}_\alpha (V) -
  \sum_\alpha   c_\alpha {\cal T}_\alpha (U_{\rm min}(V)) - T(U_{\rm
    min}(V),V) \, \right]^2 \;\;\;,
\label{eq:fit1}
\eeq
where $c_\alpha$ stands for $c_i(C)$ and ${\cal T}_\alpha (U)$ for $1/N \:
[N-{\rm Re\;Tr}(U_C)]^i$, being $\alpha$ a collective index for both the
loop topology index $C$ and the power index $i$ in Eq.~(\ref{eq:par}).\newline
The second condition is that the exact values of the FP action on the
minimizing configurations $\{U_{\rm min}\}$ are reproduced by the parametrized
FP action
\beq
{\cal A}^{\rm FP}_{\rm par} (U_{\rm min}) = {\cal A}^{\rm FP} (U_{\rm min})
\;\;\;;
\eeq
the r.h.s. is known only in approximate sense (for the instanton
configurations, we used the plaquette + $(1\times2)$ loop tree-level Symanzik
improved action). The quantity to be minimized is in this case
\beq
Q_2(c_\alpha) = \sum_{\{V\}} \left[ \sum_\alpha c_\alpha {\cal T}_\alpha
  (U_{\rm min}(V)) - {\cal A}^{\rm FP} (U_{\rm min}(V)) \, \right]^2 \;\;\;.
\label{eq:fit2}
\eeq

The summation in Eqs.~(\ref{eq:fit1}) and~(\ref{eq:fit2}) is intended to
run on a representative enough set of configurations.
We have considered a set of $\sim 500$ thermal equilibrium
configurations\footnote{The same used in Ref.~\cite{BN96}; we thank the
  authors of Ref.~\cite{BN96} for having supplied these configurations to us.}
and, in addition, the instanton configurations $\{U_3\}$ built in the previous
Section, retaining the freedom to give different weights to the first and to
the second conditions of the fit and to thermal and instanton configurations.
In formulae, the quantity to be minimized has been taken of the form
\bea
Q_{\rm tot} & = & \lambda^{\rm (therm)} [ \;
              \lambda_1^{\rm (therm)} Q_1^{\rm (therm)}
            + \lambda_2^{\rm (therm)} Q_2^{\rm (therm)} \; ] \nonumber \\
            & + & \lambda^{\rm (inst)} [ \;
              \lambda_1^{\rm (inst)} Q_1^{\rm (inst)}
            + \lambda_2^{\rm (inst)} Q_2^{\rm (inst)} \; ] \;\;\;.
\label{eq:fit}
\eea
The $Q$'s associated to each sector of the fit are polynomials quadratic
in the coefficients $c_\alpha$; the quadratic term of each polynomial
is characterized by a matrix of order equal to the number of the coefficients
$c_\alpha$ entering the action parametrization (in our case 112, coming from
28 loop topologies times 4 powers). The eigenvalue analysis of these matrices
shows that they are characterized by all but one quasi-zero modes. For
example, in the case of the matrix coming out in $Q_1^{\rm (therm)}$, the
eigenvalues, normalized to give trace 1, are distributed as follows: 0.9849
the largest eigenvalue, 0.0138 the second, 0.0007 the third, and so on in
decreasing order. The presence of so many quasi-zero modes indicates that
there are many equivalent sets of couplings $c_\alpha$ satisfying with almost
the same accuracy a given sector of the fit. In view of this relatively wide
freedom, it is reasonable to expect that there exist combinations of the
couplings $c_\alpha$ which nicely fit all the requested conditions
simultaneously. Among different equivalent parametrizations, of course the
fastest for Monte Carlo simulations should be chosen.

\begin{table}[tb]
\setlength{\tabcolsep}{0.74pc}
\caption[]{Couplings of the parametrization of the type III FP action 
determined in this work (see definition (\ref{eq:par})).} 
\centering
\begin{tabular}{ccccc}
\hline
loop & $c_1$ & $c_2$ & $c_3$ & $c_4$  \\
\hline
plaquette           & --0.544089 &   1.940530 & --0.466298 &   0.016159 \\
bent rectangle      &   0.109877 & --0.114994 & --0.008653 &   0.007343 \\
twisted perimeter-8 & --0.008914 &   0.004823 &   0.015426 & --0.001432 \\
\hline
\end{tabular}
\label{tab:new_par}
\end{table}

We have proceeded in the following way: first of all, we have restricted
ourselves to all the possible combinations of no more than three topologies of
loops among those living on the $2^4$ hypercube (see Table~\ref{tab:loops}).
We have considered combinations containing always the plaquette and one or two
other loops; moreover, we have excluded parametrizations involving two
perimeter-8 loops, which are in general too slow for standard Monte Carlo
updating algorithms. For any chosen combination, we have first performed the
fit only on the thermal equilibrium configurations, that is we have imposed
$\lambda^{({\rm therm})}=1$ and $\lambda^{\rm (inst)}=0$ in $Q_{\rm tot}$. The
weights $\lambda_1^{\rm (therm)}$ and $\lambda_2^{\rm (therm)}$ have been
fixed in order that, for the set of couplings minimizing $Q_{\rm tot}$,
both $Q_1^{\rm (therm)}$ and $Q_2^{\rm (therm)}$ are as
close as possible to their absolute minima, which are realized for
$(\lambda_1^{\rm (therm)},\lambda_2^{\rm (therm)})=(1,0)$ and
$(\lambda_1^{\rm (therm)},\lambda_2^{\rm (therm)})=(0,1)$, respectively.
We have found many different
parametrizations giving equally good fits, as expected because of the many
quasi-zero modes involved in the procedure. We could reproduce in particular
the parametrization IIIa of Ref.~\cite{BN96}, which is indeed the most
economical for numerical simulations.

\begin{figure}[htb]
\begin{center}
\leavevmode
\epsfxsize=90mm
\epsfbox{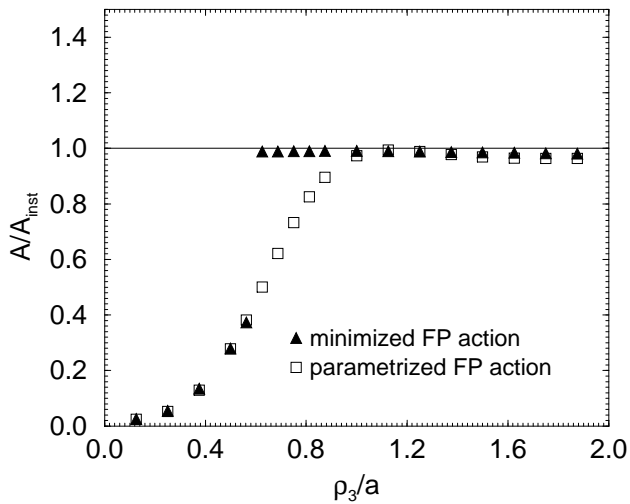}
\caption[]{As in Fig.~\ref{fig:srho_typeIII} with the minimized FP action
  (filled triangles) and the parametrized FP action defined in
 Table~\ref{tab:new_par} (squares).}
\label{fig:srho_par}
\end{center}
\end{figure}

In order to fix $\lambda_1^{\rm (inst)}$ and $\lambda_2^{\rm (inst)}$ for
the same choice of loop topologies, we have repeated the procedure
considering only the instanton configurations, that is for $\lambda^{({\rm
    therm})}=0$ and $\lambda^{\rm (inst)}=1$ in $Q_{\rm tot}$.
Then we have combined thermal sector and instanton sector of the fit, starting
with $\lambda^{\rm (therm)}=1$ and $\lambda^{\rm (inst)}=0$ and
progressively increasing the $\lambda^{\rm (inst)}$. We have realized that
for not too large values of $\lambda^{\rm (inst)}$, the quality of the fit
in the sector of the thermal configurations remains good, still a
consequence of the presence of many zero modes in this sector of the fit.
To choose among the different, equally acceptable, loop combinations, we have
measured each candidate parametrized action on the instanton configurations
$\{U_3\}$ and checked to what extent it reproduces the values of the FP action
obtained by minimization (Table~\ref{tab:min}). The last test has turned
out to be the most stringent,
leading to a strong restriction of the candidate parametrizations to a few
ones involving always the plaquette, a perimeter-6 loop (entries 2-4 in
Table~\ref{tab:loops}) and a perimeter-8 loop (entries 5-28 in
Table~\ref{tab:loops}). Once a fitting parametrization ${\cal A}_{\rm par}^{\rm
    FP}$ is found, it must be checked that it is positive (in the sense that
${\cal A}_{\rm par}^{\rm FP} \geq 0$ for any configuration), since this
property is not {\it a priori} guaranteed. We could not find any sufficient
condition for positivity which was not too restrictive and we had to turn to
numerical checks on some selected classes of configurations. Of course these
checks, although quite stringent, do not prove in absolute the positivity of
the action. We found that the positivity test represented a real bottleneck in
the selection of the parametrizations, since as soon as instantons were
introduced into the fit the candidate parametrizations lost positivity in
almost all the cases. We could escape this problem only in one case, namely
for the parametrization given in Table~\ref{tab:new_par} which involves the
plaquette, the bent rectangle (entry no.~3 in Table~\ref{tab:loops}) and the
twisted perimeter-8 loop (entry no.~25 in Table~\ref{tab:loops}). Of course,
the simulation time cost of this parametrization is high, since it involves
loops with large multiplicity and large perimeter. This action is 5 times
slower than the type IIIa parametrization of Ref.~\cite{BN96} and 35 times
slower than the Wilson action with the same Monte Carlo algorithm. In
Fig.~\ref{fig:srho_par} and in
Table~\ref{tab:min} this parametrized action is compared with the exact FP
action on the configurations $\{U_3\}$. It can be seen that the agreement is
good for instanton size $\hat\rho_3 > 0.8$. In the region $0.5<\hat\rho_3<0.8$
the quality of the fit is poor: actually we excluded from the beginning the
instanton configurations corresponding to this region, since here the Symanzik
tree-level improved action used in the minimization procedure to determine
the FP action is not completely reliable for the smallness of the involved
sizes. For $\hat\rho_3<0.5$ the agreement is again very good: since
below the critical size the blocked configurations represent small range
fluctuations around the trivial vacuum, they resemble thermal
configurations; so, the agreement in this region is actually an indication of
the quality of the fit in the sector of the thermal configurations.

A few-couplings parametrization of the FP action well reproducing the
instanton action was already found in~\cite{DeGHaKo97} using periodic b.c.
according to the technique described in~\cite{DHZ96a}.
Also in this case it was found that an additional perimeter-8 loop is
needed to well parametrize the FP action in presence
of instanton solutions, although a different choice of closed paths
turned out to be preferable, i.e. the loops no. 1, 4 and 19
of Table~\ref{tab:loops}; the computation time of this parametrization
is roughly the double of that of Table~\ref{tab:new_par},
the behavior of the action with the instanton size being
qualitatively the same.

In order to determine topological quantities from MC simulations
a further step is needed, namely a suitable definition of the topological
charge operator. The use of the FP topological charge operator $Q_{\rm FP}$ in
conjunction with the FP action allows a consistent definition of the
topology on the lattice~\cite{BBHN96,DFP97,DHZ96a,DHZ96b,DeGHaKo97},
since in this case all the configurations with non-zero topological
charge have action larger or equal than the continuum instanton action
in the given topological sector~\cite{Nie97}.
In particular, this implies that there is no over-production of
small-size instanton configurations.
This ideal picture is not automatically realized when a parametrized
version of a FP action is used in numerical simulations.
In this case it could well be that there exist configurations having
(parametrized) FP action lower than the continuum instanton action
and topological charge different from zero.
Our results (Fig.~\ref{fig:srho_par}) seem to suggest that the
FP topological charge and action reproduce the continuum values up
to size $\hat{\rho}\sim 0.5$, while the parametrization of
Table~\ref{tab:new_par} deviates from the ideal behavior at larger values
(around $0.8$). This indicates that further improvements are needed: one
possibility is to improve the parametrization itself by adding
new loops or by investigating other functional forms of the parametrized
action; another is to improve the inverse blocking procedure by increasing
the number of steps or by using a better approximation of the FP
action in the minimization procedure.

We turn now to some questions concerning the Symanzik improvement
in the context of the parametrization of a FP action. For the technical
details involved, we refer to the Appendix.

The FP action satisfies by definition the on-shell tree-level Symanzik
conditions to all order in $a^2$. For instanton configurations the Symanzik
conditions at order $a^n$ ensure the absence of the term $O(a/\rho)^n$ in
the expansion of the action in powers of $a/\rho$. Of course, a parametrized
form of the FP action does not necessarily fulfill the Symanzik
conditions. Nevertheless, a parametrized form of the FP action having exactly
vanishing cut-off effects at tree-level at the lowest order in $a^2$ is
preferable, since it would be automatically improved in the perturbative
regime and everywhere smooth configurations are physically relevant.
The on-shell tree-level Symanzik conditions at order $a^2$ would in
particular ensure that the first correction to the continuum action of
instanton solutions is $O(a/\rho)^4$; as a consequence,
the behavior of the parametrized action on instantons with large size
(not included in the fit procedure) would be improved. In
consideration of this, we have repeated the fit procedure including
with some weight the on-shell tree-level Symanzik conditions at order
$a^2$. However we were not able to find any positive
definite parametrization exhibiting a nice fit and supporting also the
Symanzik conditions.

We conclude observing that the parametrization of Table~\ref{tab:new_par}
badly violates the Symanzik conditions. This
is not in contradiction with the fact that this parametrization nicely
fits the FP action for the restricted set of configurations taken into
account in the fit procedure, since a compensation effect between
contributions of different orders in $a^2$ (equally important for the
configurations considered) occurs.

\section{Conclusions}
\label{sec:concl}

The conclusion of this work is that it is possible to carry out in a
consistent way the program of building instanton classical solutions of the FP
action of a RG transformation for the SU(3) lattice gauge theory without
imposing any b.c.. We have shown how these configurations,
defined on open lattices, can be included in the fit procedure to determine a
few-couplings parametrization of the FP action which reproduces
the scale-invariance properties of the exact FP action up to a size
$\hat{\rho}$ as small as $0.8$.
Some additional work is needed to improve the parametrization in order
to make it effective in numerical MC determinations of topological quantities.

\section{Acknowledgments}

We thank F.~Niedermayer for many useful suggestions and for critical
reading of the manuscript. We acknowledge valuable discussions with
M.~D'Elia and P.~Hasenfratz. We also acknowledge financial support
from Fondazione ``A.~Della Riccia''-Italy (F.~F.) and from INFN-Italy (A.~P.).

\appendix
\section{Appendix}

We consider lattice instanton configurations in SU(2) defined by~\cite{GGSV94}
\beq
U_\mu(x)={\rm P}\exp \int_0^a A_\mu(x+s\hat\mu) \: {\rm d}s \;\;\; ,
\label{eq:Pexp_app}
\eeq
where $A_\mu(x)=i\sum_a A^a_\mu(x) \sigma^a/2$ are the anti-Hermitian gauge
fields of the infinite volume continuum instanton solution ($\sigma^a$,
$a=1,2,3$, are the Pauli matrices). They are given by
\beq
A_\mu(x) = -i\f{\eta^a_{\mu\nu}x_\nu \sigma^a}{(x^2+\rho^2)} \;\;\; ,
\label{eq:A}
\eeq
where $\eta^a_{\mu\nu}$, $a=1,2,3$, is the 't Hooft tensor~\cite{Hoo76},
$\rho$ is the instanton size and we have assumed that the instanton is
centered at the origin. The field tensor
$F_{\mu\nu} =
\de_\mu A_\nu - \de_\nu A_\mu  + [A_\mu,A_\nu] $ turns out to be
\beq
F_{\mu\nu}(x) = 2i\eta^a_{\mu\nu}\sigma^a \f{\rho^2}{(x^2+\rho^2)^2} \;\;\; .
\label{eq:F}
\eeq

A lattice action ${\cal A}=\sum_x {\cal A}(x)$ defined in terms of the
symmetrized action density
\beq
{\cal A}(x) = \f{1}{N} \sum_{C\ni x, i\geq1} c_i(C) \f{[N-{\rm
Re\;Tr}(U_C)]^i}{{\rm
  perimeter}(C)} \;\;\;.
\label{eq:act_dens_app}
\eeq
can be expanded by using~(\ref{eq:Pexp_app}) as ${\cal A}=\sum^\infty_{n=0}
a^{4+2n} {\cal O}^{(4+2n)}$, where ${\cal O}^{(4+2n)}$ is a combination of all
the
independent gauge-invariant operators with na\"{\i}ve dimension $4+2n$. The
coefficients of this combination depend on the coefficients $c_i(C)$
for a given choice of the topology of the loops $C$ involved in the
definition of the action. The expansion of the action up to the order $a^6$ is
\bea
{\cal A}\; =\;  a^4 \sum_C r_0(C) c_1(C) \sum_x {\cal O}_0(x) + a^6 \left[
\sum_C
  r_1(C)c_1(C) \sum_x {\cal O}_1(x) \right. \nonumber \\
  +  \left. \sum_C r_2(C)c_1(C) \sum_x {\cal O}_2(x) + \sum_C r_3(C)c_1(C)
    \sum_x {\cal O}_3(x) \right] + O(a^8) \;\;\;,
\label{eq:exp}
\eea
with
\bea
{\cal O}_0(x) & = & -\f{1}{2} \sum_{\mu,\nu} {\rm
  Tr}(F^2_{\mu\nu}(x))\;\;\;, \\
{\cal O}_1(x) & = & \f{1}{12} \sum_{\mu,\nu} {\rm Tr}({\cal D}_\mu
F_{\mu\nu}(x))^2\;\;\;, \\
{\cal O}_2(x) & = & \f{1}{12} \sum_{\mu,\nu,\lambda} {\rm Tr}({\cal D}_\mu
F_{\nu\lambda}(x))^2\;\;\;, \\
{\cal O}_3(x) & = & \f{1}{12}\sum_{\mu,\nu,\lambda} {\rm Tr}({\cal
  D}_\mu F_{\mu\lambda}(x){\cal D}_\nu F_{\nu\lambda}(x))\;\;\;,
\eea
being
${\cal D}_\mu F_{\nu\lambda} = \de_\mu F_{\nu\lambda}
+[A_\mu,F_{\nu\lambda}]$. The coefficients $r_0$, $r_1$,
$r_2$ and $r_3$ are given in Table~\ref{tab:loops} for all the loops which
live on a hypercube $2^4$. They have been found by considering the
expansion~(\ref{eq:exp}) on configurations representing constant fields with
low intensity. The normalization condition $\sum_C r_0(C) c_1(C)=1$ has to be
satisfied in order to have the correct continuum limit. The on-shell
tree-level Symanzik improvement at $O(a^2)$ is realized by imposing $\sum_C
r_1(C)c_1(C)=0$ and $\sum_C r_2(C)c_1(C)=0$; no condition comes from the
operator ${\cal O}_3(x)$, since it is identically zero on
configurations which satisfy the equations of motion~\cite{LW85}.

On an instanton classical solution~(\ref{eq:A}), the
operator ${\cal O}_3(x)$ is identically zero, while the other operators are
\bea
{\cal O}_{0,{\rm inst}}(x)&=&\f{48\rho^4}{(x^2+\rho^2)^4}\;\;\;,\nonumber \\
{\cal O}_{1,{\rm inst}}(x)&=&\f{-48\rho^4x^2}{(x^2+\rho^2)^6}\;\;\;,
\label{eq:op_inst} \\
{\cal O}_{2,{\rm inst}}(x)&=&\f{-192\rho^4x^2}{(x^2+\rho^2)^6}\;\;\;.\nonumber
\eea
We observe that the operators ${\cal O}_1(x)$ and ${\cal O}_2(x)$ are not
independent when acting on the instanton configurations. On such
configurations, the $O(a^2)$ term in the action vanishes if the
weaker condition $\sum_C (r_1(C) + 4 r_2(C)) c_1(C) =0$ is satisfied.
The contribution of the operators ${\cal O}_1(x)$ and ${\cal O}_2(x)$ to the
action~(\ref{eq:exp}) inside a hypersphere with radius $R$ centered in the
origin can be estimated by integrating the expressions on the r.h.s. of
Eqs.~(\ref{eq:op_inst}) on the continuum. It turns out that
\bea
\int_{|x|<R} \: d^4x \; {\cal O}_{0,{\rm inst}} & = & 8\pi^2\left[ 1 -
  \f{1+3(R/\rho)^2}{[(R/\rho)^2+1]^3}\right] \nonumber \\
& \stackrel{\rho\ll R}{\longrightarrow} & 8\pi^2\left[1 -
  3\left(\f{\rho}{R}\right)^4 +
O\left(\f{\rho}{R}\right)^6 \right] \nonumber \\
a^2 \int_{|x|<R} \: d^4x \; {\cal O}_{1,{\rm inst}} & = &
8\pi^2\left(-\f{1}{5}\right)\left(\f{a}{\rho}\right)^2 \left[1-\f{10
    (R/\rho)^4+5(R/\rho)^2+1}{[(R/\rho)^2+1]^5} \right] \nonumber \\
& \stackrel{\rho\ll R}{\longrightarrow} &
8\pi^2\left(-\f{1}{5}\right)\left(\f{a}{\rho}\right)^2
\left[1-10 \left(\f{\rho}{R}\right)^6 + O\left(\f{\rho}{R}\right)^8\right]
\nonumber \\
a^2 \int_{|x|<R} \: d^4x \; {\cal O}_{2,{\rm inst}} & = & 4 \; a^2
\int_{|x|<R}  \: d^4x \; {\cal O}_{1,{\rm inst}} \label{eq:int}
\eea
It can be seen from these results that the integration of an operator
with na\"{\i}ve dimension $4+2n$ outside a region with characteristic size $R$
in the Euclidean four-dimensional space scales as
$(a/\rho)^{2n}(\rho/R)^{4+2n}$.

Taking the infinite volume limit in Eqs.~(\ref{eq:int}), the lattice action of
an instanton with size $\rho$ can be estimated as
\beq
{\cal A} = 8\pi^2\left[1 - \f{1}{5}\left(\f{a}{\rho}\right)^2 \sum_C ( \;
    r_1(C) + 4 r_2(C)\; )\;c_1(C) +
    O\left(\f{a}{\rho}\right)^4 \right] \;\;\; .
\label{eq:exp_inst}
\eeq
For the Wilson action, the plaquette + $(1\times2)$ loop tree-level Symanzik
improved action ($c_1({\rm plaq})=5/3$, $c_1(1\times2)=-1/12$) and the
plaquette + $(2\times2)$ loop tree-level Symanzik improved action ($c_1({\rm
  plaq})=4/3$, $c_1(2\times2)=-1/48$), also the terms $O(a/\rho)^4$ of the
expansion~(\ref{eq:exp_inst}) are known~\cite{GGSV94}:
\bea
{\cal A}_{\rm Wilson} = 8\pi^2\left[1 - \f{1}{5}\left(\f{a}{\rho}\right)^2
               - \f{1}{70}\left(\f{a}{\rho}\right)^4
               + O\left(\f{a}{\rho}\right)^6 \right] \;\;\; , \\
{\cal A}_{{\rm Sym}, 1\times2} = 8\pi^2\left[1
                        - \f{17}{210}\left(\f{a}{\rho}\right)^4
                        + O\left(\f{a}{\rho}\right)^6 \right] \;\;\; , \\
{\cal A}_{{\rm Sym}, 2\times2} = 8\pi^2\left[1
                        + \f{2}{35}\left(\f{a}{\rho}\right)^4
                        + O\left(\f{a}{\rho}\right)^6 \right] \;\;\; .
\eea

\end{document}